Short Paper

# Development of Augmented Reality Application for Made-to-Order Furniture Industry in Pampanga, Philippines

Jaymark A. Yambao
College of Computing Studies, Mexico Campus, Don Honorio Ventura State University
jayambao@dhvsu.edu.ph
(corresponding author)

John Paul P. Miranda
College of Computing Studies, Mexico Campus, Don Honorio Ventura State University
jppmiranda@dhvsu.edu.ph

Earl Lawrence B. Pelayo
College of Computing Studies, Bacolor Campus, Don Honorio Ventura State University
elbpelayo@dhvsu.edu.ph

**Abstract**

*Purpose* – The focus of the study was to develop a mobile application utilizing marker-less augmented reality for specific made-to-order products to support furniture and fixtures businesses.

*Method* – The study implemented mixed-methodology to properly identify the various stakeholders' considerations in developing the application. Interviews with key informants were conducted to ensure that the features were appropriate for the intended user needs, and selected ISO standards were used as evaluation criteria.

*Results* – The results indicate that the mobile application with marker-less AR technology was found to be highly acceptable by three evaluators (i.e., customers, owners, and IT experts).

*Conclusion* – The study also highlighted the use of AR-related technology in this case, where marker-less has the potential to improve customer purchasing experience even further.

*Recommendations* – Future studies may include using newer technologies to further improve the application.

*Practical Implications* – The study suggests that Augmented Reality technology could be used to connect specific businesses directly to consumers regardless of setting or context.

## INTRODUCTION

The COVID-19 pandemic helps us appreciate the role of technology more than ever, particularly the things we take for granted, such as the way we interact with things and with one another. Rapid technological progress is a highlight during this period, particularly in multiple industries. Businesses, for example, are looking for new and existing technologies to enhance their business processes and practices. In recent years, technological innovations have also played pivotal roles and shifted paradigms in how businesses integrate and engage their customers about their services and products (Muratovski, 2015). E-commerce offers a great opportunity for any business because it delivers flexibility in terms of time and location. Customers use e-commerce applications to buy stuff online where there is a vast range of products and services (Geetha & Rathi, 2020). Also, it provides convenience and saves time since customers can buy any product from their offices, in the comfort of their homes, or anywhere in the world through the use of the internet (Rahman et al., 2018).

Augmented Reality (AR) is one of the emerging technologies for ecommerce in the 21st century (Yamin, 2019; Young & Smith, 2016). The popularity of AR on mobile devices is due to advancements in smartphone technology (Sudarshan, 2018). Also, AR has matured significantly over time, from the conceptual belief that augmented reality answers the important technical obstacles by creating applications that augment virtual objects in the real world that are entertaining and usable (Aljojo et al., 2020; Gjøsæter, 2015; Young & Smith, 2016). With the number of devices that AR is capable of, as reflected in billions of downloads of Google Play Services, AR could change the way we do business online. For example, AR can bring life to a product and provide immersive content (Berryman, 2012; Yim et al., 2017; Oh et al., 2008). It also creates a sense of presence for objects that are not really there (Jung et al., 2015; Krevelen & Poelman, 2010; Yim et al., 2017). As a result, virtual objects merged with AR affect the decisions of possible customers a lot more than the images and videos usually used in traditional e-commerce experiences (Abrar, 2018; Flavián et al., 2019). Applications with AR offer a great experience and improve customer engagement by enabling the ability to customize customers' preferences (Baytar et al., 2020; Berryman, 2012; Flavián et al., 2019; Richter & Raška, 2017; Yaoyuneyong et al., 2016; Yim et al., 2017; Oh et al., 2008) while allowing customers to explore their options while buying through an online application (Richter & Raška, 2017). Another example is that, AR-integrated advertisements are favored over traditional advertisements among potential customers (Bilgili et al., 2019).

Likewise, AR allows the improvement of the real world by placing a virtual object in real-time (Baytar et al., 2020; Carmigniani & Furht, 2011; El-Seoud & Taj-Eddin, 2019). Such innovations could also evoke higher purchase intentions (Abrar, 2018; Adam & Pecorelli, 2018; Baytar et al., 2020; Richter & Raška, 2017), provide new opportunities (Carvalho et al., 2011; El-Seoud & Taj-Eddin, 2019; Richter & Raška, 2017) for businesses to reach out to customers (Carmigniani & Furht, 2011), develop possibilities for the customer's engagement (Abrar, 2018; Baytar et al., 2020; El-Seoud & Taj-Eddin, 2019; Yaoyuneyong et al., 2016), add value to their products (Flavián et al., 2019), and stay ahead of the competition (Abrar, 2018; El-Seoud & Taj-Eddin, 2019; Oh et al., 2004). In addition, Lu and Smith (2008) said that AR can be used to provide more accurate information on products like furniture, jewelry, accessories, and other decorative products that need space and use volume (Carvalho et al., 2011; Oh et al., 2008). AR technology also provides the user the power to make better decisions when purchasing these kinds of products. Accenture discovered in a more recent survey on the use of AR in e-commerce that 88% of customers indicated that the use of AR in furniture businesses will likely increase their intention to purchase a furniture product (Accenture, 2014). This was supported by Flavián et al. (2019), which said that by incorporating AR technology, businesses can optimize the user experience for purchasing things (Oh et al., 2004; Oh et al., 2008). For example, AR technology can enhance the customer's buying experience in pre-purchase scenarios, wherein the customer may view a product in real-time located in their chosen places based on the color of the paint on their walls and ceiling and other decorations in their house. As mentioned by Baytar et al. (2020), AR could be used to improve the online shopping experience, which can lead to higher revenues (Bucko et al., 2018).

This prompts the study to develop a mobile application with a product preview using AR for made-to-order products. It aims to provide a platform for furniture and fixtures businesses to promote their products by giving customers the power to personally feel and evaluate the product at their own time and place, which in turn enhances customers' engagement and buying experience. Specifically, the study focused on 1) designing and developing a website; 2) developing and integrating a mobile-based application using AR technology specifically for furniture and fixture businesses; and 3) evaluating the compliance of the mobile application to International Organization for Standardization (ISO) 25010.

# METHODOLOGY

The study used a mixed-method for identifying the design considerations of the AR mobile application. The study adopted the contextual design approach. This method was followed to ensure that it supports and meets the needs of the intended users (i.e., furniture business owners and potential customers) (Holtzblatt & Beyer, 1997). In the qualitative part of the study, an interview with three informants (i.e., production staff, secretary, and store manager) was initiated. These informants were selected due to their experience, involvement, and existing knowledge of the made-to-order furniture business. The informants were asked about the existing processes regarding their ordering procedures and the issues they usually faced during inquiries related to the made-to-order products. Questions related to the possible functionalities that might be helpful to their business were also asked. All the interviews lasted about 10 to 30 minutes. A clarification interview was also conducted during the development process in order to ensure that the application caters to the needs of the intended users. Video recordings were utilized with the proper consent of the informants to document all of their responses.

For the quantitative study, the developed application was evaluated by 35 customers (i.e., 10 potential and 15 existing customers), five staff, and five IT experts using a 4-point Likert scale (4 being the highest) and selected ISO 25010 quality standards (i.e., Functional Suitability, Usability, and Portability) (see Table 1). Responses from the questionnaire are coded as "strongly disagree", "disagree", "agree" and "strongly agree". A verbal interpretation ranged from 4 as highly acceptable to 1 as not acceptable. The evaluation was purposive in nature and was conducted in Pampanga in different locations over one week. The overall mean rating of the evaluations was then reported.

Table 1. Selected ISO 25010 Quality Standard Evaluation Questionnaire

| Criteria | Indicators |
|---|---|
| **Functional Suitability** | Usefulness and appropriateness |
| | Real-time preview |
| | Product traceability |
| | Absence of failures |
| **Usability** | Ease of execute |
| | User-friendly |
| | Provide updates |
| | Order management |
| **Portability** | Error-free installation |
| | Adaptability |
| | Availability |

# RESULTS AND DISCUSSION

## *Development of the E-commerce Website*

Based on the series of interviews among the key informants, the website should have the following main features: 1) display available furniture and services; 2) improve customer interaction; and 3) provide automation for processing customer orders and furniture delivery. Figure 1 showcases the main categories of furniture, namely: living room, dining set, garden set furniture, and day bed. According to the informants, these four (4) furniture types are their main products.

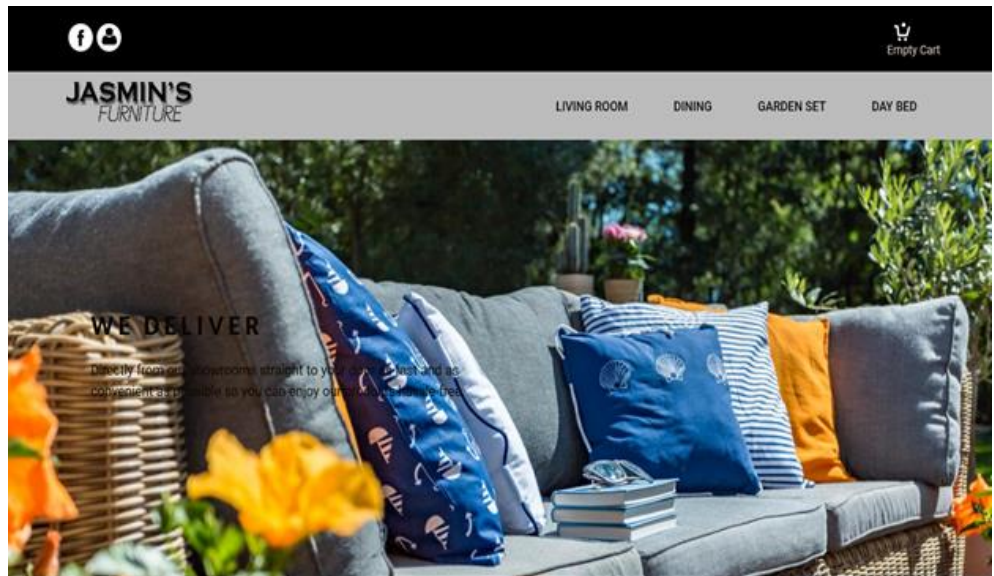

*Figure 1.* Home page

According to the informants, the website should also be able to showcase featured furniture (Figure 2). This way, the future customer will be able to see what kind of furniture the business usually creates. The website also has the ability to purchase products online, track the production status of made-to-order furniture, and customize and personalize furniture before purchase. This is consistent with the study of Lu and Smith (2008) that found that these kinds of features would greatly increase the customer's comfort in buying and purchasing furniture online.

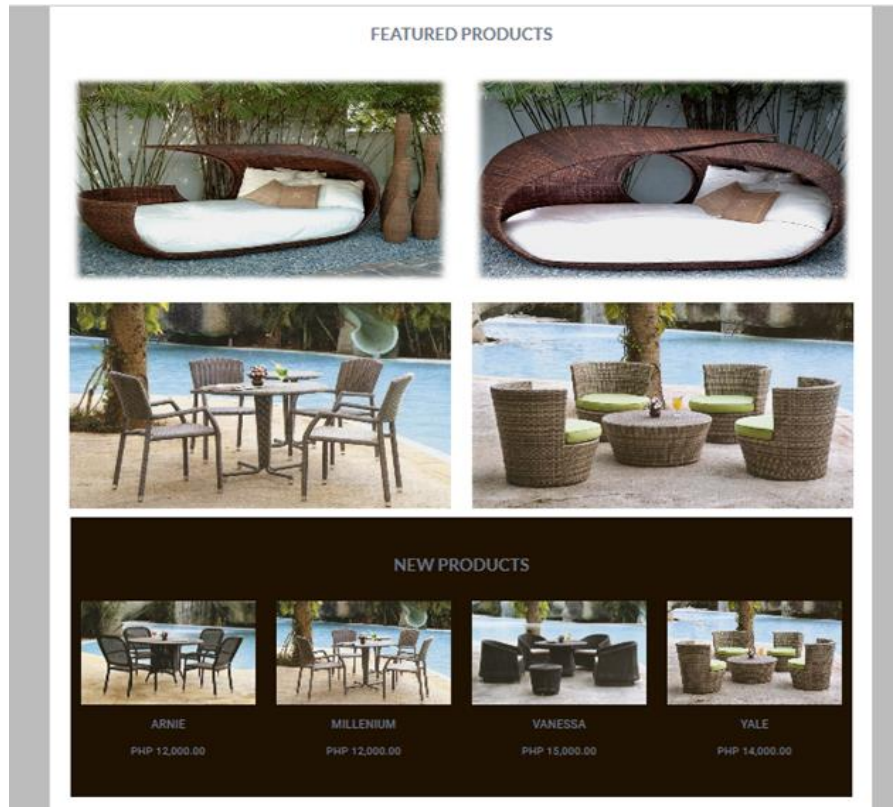

*Figure 2.* Featured Products and New Products

***Development of Mobile Application and Integration of AR Technology***

One of the main problems on an e-commerce website was that the videos and pictures that advertise certain products like furniture are different from the actual product. Specifically, size and color that complement the interior of the house. This warranted the study to use AR technology to help both the customer and the furniture business owner ease the problem mentioned. Suggestions based on the interviews related to these served as a guide in the development of the mobile application with AR technology. The prototype was developed in Unity 3D, while the AR technology integration utilizes the Vuforia engine. The 3D drawings initially used in the prototype were provided by the informants and recreated and modeled using the SketchUp software. This integration paved the way for the application to be capable of previewing the products directly to the desired location using the actual sizes and designs before purchase. According to the informants, this feature is needed as it can help both the customer and the business owner visualize the desired product, thus enhancing the customer's buying experience (Figure 3).

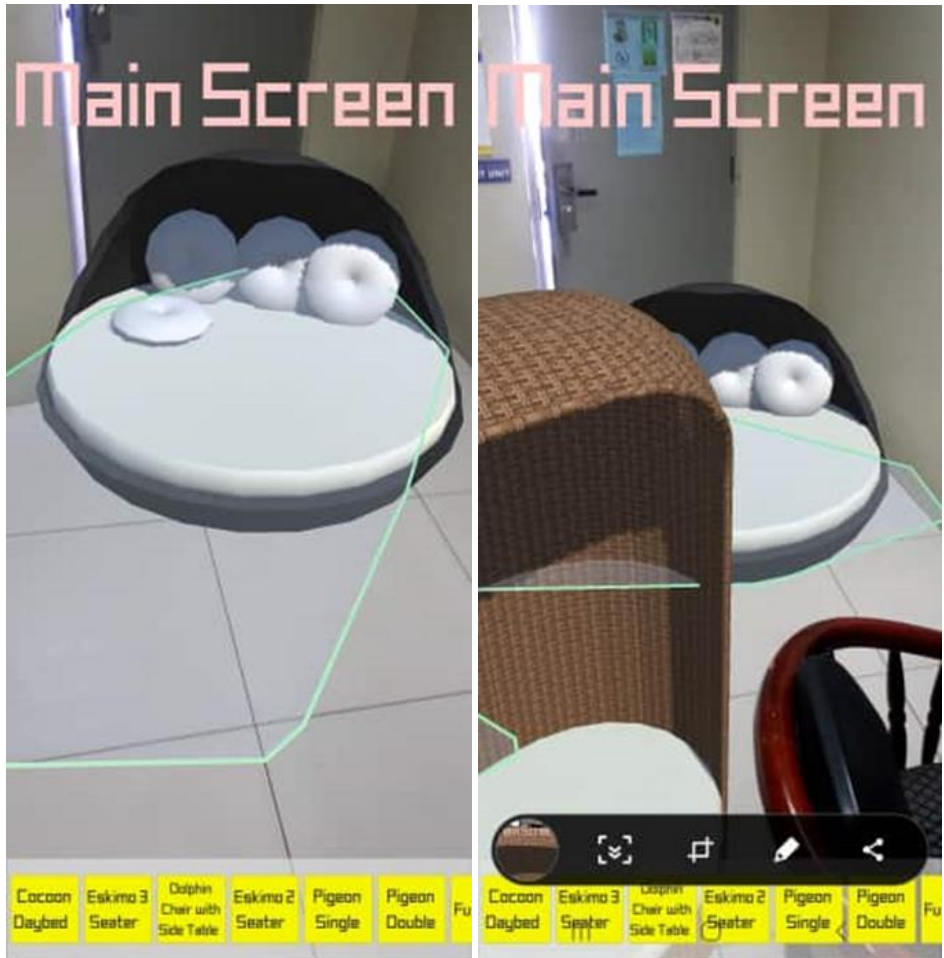
*Figure 3.* Screenshots of the mobile AR application

*Prototype Evaluation*

Table 2 shows that the application is rated as highly acceptable among the respondents of the study. This rating was observed in several aspects of the ISO 25010 quality standards. For functionality, the product was able to provide real-time experiences in terms of the product position (Mean = 3.84, n = 35) and scale (Mean = 3.84, n = 35), which was in previewing the actual product. In terms of its product preview feature, the application was also rated as being consistent with real-world (Mean = 3.80, n = 35). Additionally, the ability to monitor the product status and development was evaluated as highly acceptable (Mean = 4.0, n = 35). This indicates that the AR mobile application provides a real-time experience for both customers and business owners, making the process of purchase and visualization seamless in terms of customer engagement and buying experience. For usability, both the website and the application were user-friendly (Mean = 3.72, n = 35), provided clear instruction, particularly in terms of its navigation features (Mean = 3.76, n = 35), and the ability to showcase and preview the product in a desired location and track production status. The usability of the application was evaluated as highly acceptable in terms of its ability to visualize products in real-time that improve the buying experience for both the customer and the business owner. For portability, the application can easily be moved to another location and can be accessed anytime using an internet connection. This indicates that the application was readily available anytime and anywhere whenever the customer felt the need to evaluate the product inside the house, in the garden, or any other place where they desired.

Table 2. Overall Selected ISO 25010 Survey Results

| Software Standards | Mean (M) | | | Verbal Description |
|---|---|---|---|---|
| | Customer | Business | IT Expert | |
| Functional Suitability | 3.81 | 3.80 | 4.00 | Highly Acceptable |
| Usability | 3.76 | 3.33 | 3.93 | Highly Acceptable |
| Portability | 3.78 | 3.40 | 4.00 | Highly Acceptable |
| Overall Mean | 3.80 | 3.61 | 3.91 | Highly Acceptable |

*Legend: 3.26 – 4.0 (Highly Acceptable); 2.51 – 3.25 (Moderately Acceptable); 1.76 – 2.50 (Acceptable); 1.0 – 1.75 (Not Acceptable)*

## CONCLUSIONS AND RECOMMENDATIONS

The mobile application was evaluated as highly acceptable across three selected ISO 25010 standards. It showed that AR technology provides a way for businesses to enhance their customers' buying experiences. The study has also shown that real-time status updates and product previews based on the desired location of the made-to-order product were the most sought-after features in the furniture business. The study also highlighted the use of AR-related technology in this case, which has the capability to further improve the customer buying experience. Future studies may include using newer technologies (e.g., mixed reality) to improve the application, as well as using the same technology on other business entities that require product visualization.

## PRACTICAL IMPLICATIONS

The study showed that AR-related technologies could be helpful to local industries in the Philippines. It also shows that such technology can be used by specific business models like made-to-order furniture, where instantaneous replies and spatial information are needed. Moreover, the study also shows that businesses, regardless of context, can connect with their potential consumers using emerging technologies like AR.

# ACKNOWLEDGEMENT

The authors are indebted to Don Honorio Ventura State University for funding this study.